\documentclass[aps,preprint,tightenlines,amsmath]{revtex4}

\usepackage{graphicx}
\usepackage{dcolumn}
\usepackage{bm}

\begin{document}


\title{Application of Coarse Integration to Bacterial Chemotaxis}

\author{S.~Setayeshgar\footnote
{Corresponding author.  E-mail: simas@princeton.edu, 
Tel: (609) 258-4320, Fax: (609) 258-1549},
}
\affiliation{Department of Physics, Princeton University, Princeton, NJ 08544}

\author{C.~W.~Gear,}
\affiliation{NEC Research Institute, 4 Independence Way, Princeton, NJ 08540,
and Department of Chemical Engineering, Princeton University, Princeton, NJ 08544}

\author{H.~G.~Othmer,}
\affiliation{Department of Mathematics, University of Minnesota, Minneapolis, MN 55455}

\author{I.~G.~Kevrekidis}
\affiliation{Department of Chemical Engineering, and Program in 
Applied and Computational Mathematics, Princeton University, 
Princeton, NJ 08544}

\vspace{12pt}
\date{\today}

\vspace{12pt}
\begin{abstract}
We have developed and implemented a numerical evolution scheme for
a class of stochastic problems in which the temporal evolution occurs
on widely-separated time scales, and for which the slow evolution can
be described in terms of a small number of moments of an underlying
probability distribution.  We demonstrate this method via a numerical
simulation of chemotaxis in a population of motile, independent
bacteria swimming in a prescribed gradient of a chemoattractant. The
microscopic stochastic model, which is simulated using a Monte Carlo
method, uses a simplified deterministic model for
excitation/adaptation in signal transduction, coupled to a realistic,
stochastic description of the flagellar motor.  We show that
projective time integration of ``coarse'' variables can be carried
out on time scales long compared to that of the microscopic dynamics.
Our coarse description is based on the spatial cell density
distribution.  Thus we are assuming that the system ``closes'' on this
variable so that it can be described on long time scales solely by the 
spatial cell density.  Computationally the variables are the components 
of the density distribution expressed in terms of a few 
basis functions, given by the singular vectors of the spatial density 
distribution obtained from a sample Monte Carlo 
time evolution of the system.  We present numerical results and 
analysis of errors in support of the efficacy of this 
time-integration scheme.
\end{abstract}

\maketitle

\section{Introduction}

A recurring bottleneck in computational modeling of physical processes
is the existence of multiple space and/or time scales. Important
examples include patterns in fluids, defect dynamics in solids, and
molecular dynamics of macromolecules.  For example, in the latter case, the
evolution of the physical configuration from initial to final state
typically occurs on the time scale of milliseconds, while the
interactions between constituent components must be resolved on the
picosecond time scale.  Furthermore, it is increasingly important in
simulating biological systems to have a hybrid computational framework
for interacting stochastic and deterministic processes, which occur on
different time scales.  Multiscale computational methods combine
processing on fine space/time scales according to the governing
microscopic description with macroscopic changes on a coarse grid
(see for example, \cite{ref:Shenoy,ref:Garcia,ref:Kevghkrt}).

Here we present a computational scheme for ``coarse projective'' time
integration of the macroscropic dynamics of stochastic processes that
involve multiple, widely-separated time scales.  To illustrate the
method we apply it to a microscopic model that generates the biased
random walk describing the macroscopic motion of a bacterium such
as {\em E. coli} in an attractant gradient.  Details of this
Monte Carlo model of bacterial chemotaxis that integrates signal
transduction with the motor response are given in Section \ref{sec:MCmodel}.
The focus of our analysis in this paper is the computational scheme,
and as such, our results address its plausibility rather than the
phenomenology of chemotaxis.

In Section\ \ref{sec:previous} we discuss previous work on coarse time
integration of microscopic dynamics and the new features of the method
developed here, and we motivate its application to chemotaxis.  The
Monte Carlo description is presented in Section\
\ref{sec:chemotaxis}. We outline the coarse integration scheme in
Section\ \ref{sec:ci}, and present numerical results and analysis of
errors in Section\ \ref{sec:results}.  We conclude with future
extensions and applications of this work.

\section{Previous work}
\label{sec:previous}

The traditional approach to studying long-term dynamics of multiscale
processes involves (a) the derivation of a ``coarse-grained'' set of
evolution equations, followed by, (b) the analytical and/or
computational study of these reduced equations using established,
continuum numerical analysis tools.  Recently, a so-called
``equation-free'' approach to the study of the coarse-grained behavior
of such problems has been proposed which circumvents the first
step (\cite{ref:TQK,ref:GKT,ref:LKGK}).  
This computational approach is based on the ``coarse," or
macroscopic time-stepper, a map from the coarse variables at time $t =
0$ to those at time $t = T$, where $T$ is typically much
larger than characteristic microscopic time scales in the system.  
This map is not obtained directly from
the macroscopic evolution equations, which we may not know,
but rather through short time evolution intervals of appropriately initialized
microscopic simulations.  The initial macroscopic variables are {\em
lifted} to microscopic variables to initialize the microscopic
simulation.  At the completion of a burst of microscopic simulation,
the microscopic variables are {\em restricted} back to macroscopic
variables, providing an approximation to the macroscopic time step.
This provides a {\em chord} of the macroscopic solution, which is an
approximation to the time derivative of the macroscopic solution.  This
value can then be used in any conventional continuum numerical method
for the macroscopic equations.  This approach has been applied in
several microscopic contexts, and the results
appear to be promising \cite{ref:MMK,ref:MMPK,ref:SGK,ref:HK}.

In this work we apply the coarse timestepper in a projective
integration study of a spatially-distributed kinetic Monte Carlo
simulation of a biased random walk.  When, as in this case, the
microscopic equations are stochastic, an effective algorithm must
reduce the variance inherent in individual realizations of
the stochastic process to a level that can be
tolerated by the continuum numerical algorithm applied to the coarse
system.  This can be done by lifting to multiple copies, or, as is
done here, by estimating the derivative from a least-squares fit to a
large number of microscopic time steps.  The projective integration
method applied here uses a derivative estimate for the projective step,
so that the stochastic noise is amplified by the inverse of the
effective step length used in the derivative estimate.  Hence,
variance reduction is very important.  
Augmenting the number of copies of the simulation (which, for our
noninteracting particle model corresponds also to a simulation with a
larger number of cells) is the most direct approach to variance
reduction; other variance reduction schemes are discussed in
\cite{ref:Oettinger1,ref:Oettinger2}.

In the context of bacterial chemotaxis, the computational scheme developed here can
be viewed as a direct method to compute the macroscopic evolution of
the cell density in space and time without actually deriving these
equations. An alternate approach begins with the transport equation
for the velocity jump process, in which discontinuous changes in
the direction (or speed) of an individual cell are generated
by a Poisson process.  It can be shown rigorously that this
reduces to the chemotaxis equation under suitable scaling of
space and time \cite{ref:Othmer:DLT:2002}
\begin{equation} 
\label{chemo-eqn} 
\dfrac{\partial \mu_0}{\partial t} =
\nabla \cdot(D \nabla \mu_0 - \mu_0 \chi(S)\nabla S) ,
\end{equation} 
In the above, $\mu_0(x,t)$ represents the density of particles at spatial
position $x$ at time $t$, $D$ is the diffusion constant, 
$S$ is the concentration of the chemotactic 
attractant/repellent, and $\chi(S)$ is the chemotactic sensitivity.
However, there is as yet no analytical procedure available 
for determining the diffusion constant and chemotactic sensitivity
when dependence on internal state variables determining the cell's 
response to the external signal is explicitly included the
transport equation.  We show that a macroscopic time-stepper can
be used here in the time evolution of the spatial density, 
even though the macroscopic evolution equations are
not known.

\section{Bacterial chemotaxis} \label{sec:chemotaxis}
\label{sec:MCmodel}
\subsection{Signal transduction}

The ability to sense and respond to  environmental cues is necessary
for the survival of most organisms. 
\textsl{Escherichia Coli} ({\em E. coli}) is a common and well-known single
cell organism, with roughly $4000$ genes, whose chemotaxis network 
has emerged as a prototype for understanding signal transduction networks
in general \cite{ref:BactRev}.  Its genome is known, the crystal structures
of many proteins have been obtained, and a large number
of mutant strains exist, allowing detailed behavioral studies. 

For each cell, chemotactic behavior begins when attractant or
repellent molecules bind to membrane receptors, triggering a cascade of chemical
reactions inside the cell that culminates in the production of the phosphorylated
form of a response regulator protein (CheY-P) which 
controls the direction of rotation of the flagellar motor.
The series of reactions that converts the extracellular signal
(attractrant/repellent) into cellular response is referred to as the
signal transduction pathway.
Flagella possess an inherent chirality, such that counter-clockwise (CCW)
rotation results in bundling-up of the six to eight flagella per
cell, allowing them to act as a single propeller and leading to smooth
swimming motion of the bacterium.  When the flagella rotate
clockwise (CW), the bundle flies apart and the bacterium tumbles.
As conditions become increasingly favorable due to
increase in chemoattractant concentration, a bacterium 
extends its run-length; otherwise, 
it tumbles, and the subsequent direction 
of motion is randomly chosen, allowing a more 
favorable direction to be discovered.
The resulting motion is a biased random walk
toward favorable conditions and away from 
less favorable ones.  

The sequence of biochemical reactions that take place inside
a bacterium, starting with the binding of
an attractant or repellent molecule to receptors on the
cell surface and leading to the change in concentration of the response 
regulator species, CheY-P, has been extensively studied.
Both deterministic models of these biochemical pathways,
based on the law of mass action and 
Michaelis-Menten kinetics \cite{ref:Barkai_Leibler,ref:Spiro_Othmer}, 
as well as fully stochastic models exist \cite{ref:Bray_98}.  
Measurement of concentration change is achieved
through a temporal rather than a spatial comparison:  fast sampling of the present
external concentration is compared with memory of that concentration
some time ago.  Order-of-magnitude analyses for why
measurement of concentration changes as spatial gradients across the cell length
is not physically feasible have been given \cite{ref:Berg_Purcell}.

Memory in the signal transduction network is achieved through the existence of
fast and slow reaction time scales.  The fast reactions are receptor-ligand binding
and phosphorylation kinetics; the slow reactions
are methylation and demethylation.  Ligand binding reduces 
the autophosphorylation rate of the corresponding membrane-bound
receptor, in turn decreasing the rate of transfer of phosphoryl groups to CheY,
and resulting in a (fast) drop in [CheY-P].
Addition of methyl groups to
ligand-bound receptors restores the autophosphorylation rate, resulting in (slow)
increase in [CheY-P].  The rate of demethylation becomes significant once
a high methylation level is achieved.  Hence, [CheY-P] reflects the balance between
the fraction of ligand-bound receptors and methylated receptors.  
Perfect adaptation refers to the return of [CheY-P] to the
same steady state level, regardless of the constant concentration level of the
external stimulus.  This value falls within the fixed operational range 
of the motor response to CheY-P.

A minimal model representing the signal transduction process 
includes fast excitation and slow adaptation
to an external stimulus and is given in \cite{ref:Othmer_Schaap}:
\begin{eqnarray}
\label{eq:two_variable_model1}
	\frac{d u_1}{d t} & = & \frac{\left( f(S) - u_2 \right) - u_1}{\tau_e}, \\
\label{eq:two_variable_model2}
	\frac{d u_2}{d t} & = & \frac{f(S) - u_2}{\tau_a},
\end{eqnarray}
where $\tau_e$ and $\tau_a$ are the excitation and adaptation times, respectively,
with $\tau_e \ll \tau_a$.  We identify $u_1$ with the deviation of [CheY-P] from
its steady state value, and $u_2$ as the number of methylated receptors per unit volume.
$f(S) = f(S(t))$ is a function of the external stimulus; for example,
it is the number of bound receptors per unit volume in the presence of 
an external signal concentration, $S$:
\begin{equation}
	f = N_T S/(K_L + S).
\end{equation}
$N_T$ is the total number of receptors per unit volume, and $K_L$ is the dissociation 
constant for the ligand-binding process. 

The formal solution to Eqs.\ (\ref{eq:two_variable_model1})--(\ref{eq:two_variable_model2}) 
can be easily obtained for each cell as
\begin{eqnarray}
	u_2(t) & = & \frac{e^{-t/\tau_a}}{\tau_a} \int^t f\left(S(t')\right) 
                                 e^{t'/\tau_a} d t', \\
	u_1(t) & = & \frac{e^{-t/\tau_e}}{\tau_e} \left(\int^t f\left(S(t')\right) 
                                 e^{t'/\tau_e} d t'
	            - \frac{1}{\tau_a} \, \int^t e^{-t' \left(1/\tau_a - 1/\tau_e\right)} 
                      \int^{t'} f\left(S(t'')\right) e^{t''/\tau_a} 
                      d t'' \, d t' \right).
\end{eqnarray}
However, because $S(t) = S(x(t),t)$, where $x(t)$ is 
a random variable that represents the cell's
position at time $t$ in a given external concentration field, the
integration takes place along the cell trajectory, which is a biased
random walk: $x(t^\prime)$ for $t^\prime < t$ biases the probability of $x(t)$.
Hence, $S(t)$ is a stochastic variable, and the integration must
be carried out computationally.

\subsection{Motor and cell response}

We adopt a stochastic approach to modeling the response of
the flagellar motor to the regulator species, CheY-P.
The motor is assumed to be in one of
two states, CW or CCW, corresponding to clockwise or counterclockwise
rotation \cite{ref:Block_83,ref:Macnab}.  
The rates of transition between these states are functions
of [CheY-P]:
\begin{equation}	
CCW_i {\,\,\,\,}^{\stackrel{ k_+}{\longrightarrow}}_
                         {\stackrel{\longleftarrow}{ k_-}} \,\,\,\,
CW_i \qquad i=1,..., N_f \, , 
\end{equation}
where $N_f$ is the number of flagella per cell.
For any concentration of CheY-P, the motor has a nonzero
probability of being in either state.  The probability of waiting at least
a time $T$ for a transition from CCW to CW, $\Phi_+$, or a
transition from CW to CCW, $\Phi_-$, is given by:
\begin{equation}
	\Phi_\pm = e^{-\int_0^T k_\pm(t) dt} \, .
\end{equation}
If the motor is probed at a time $T \equiv \Delta t$, 
such that $k_\pm \Delta t \ll 1$, then the probability 
that a change in direction will occur in $(T, T+\Delta t)$
can be approximated as
\begin{equation}
\label{eq:prob_switch}
	1- \Phi_\pm \approx k_\pm(0) \Delta t \, .
\end{equation}
Systematic studies have been undertaken
that show simulation results are qualitatively 
independent of the choice of $\Delta t$
once the above restrictions are met. 

In this work, we use equilibrium transition rates, $k_\pm$, 
consistent with recent experimental measurements \cite{ref:Cluzel} of the 
single flagellum clockwise bias, $b_{CW}$, and the reversal frequency, $w$,
as functions of [CheY-P].  The rates are related to these measured quantities
according to \cite{ref:Scharf_Turner}:
\begin{eqnarray}
	b_{CW} & = & \frac{k_+}{k_+ + k_-} \, , \\
	w      & = & \frac{2 k_+\, k_-}{k_+ + k_-} \, ,
\end{eqnarray} 
where the reversal frequency is the geometric mean of the transition rates.
We take $b_{CW}$ to be described by a Hill function, $b_{CW} = {y_p}^h/({K_M}^h +  {y_p}^h)$,
where $y_p = {\rm [CheY-P]}$, with experimentally measured values for the 
Hill coefficient and dissociation constant, equal to $h=10.3$ and 
$K_M = 3.1 \, \mu {\rm M}$.  
In these experiments, it was noted that the data were consistent with 
$w = (2 \, \mu {\rm M} \, s^{-1}) \, \partial b_{CW}/\partial y_p$.  
Based on these results, we use
\begin{eqnarray}
	k_+ & = & \frac{h \, {y_p}^{h-1}}{{K_M}^h + {y_p}^h} \, , \\
	k_- & = & \frac{1}{y_p} \, \frac{h \, {K_M}^h}{{K_M}^h + {y_p}^h} \, .
\end{eqnarray}

In a constant chemoattractant field, \cite{ref:Ishihara_83,ref:Spiro_Othmer} show that
the ``Voting Hypothesis'' is successful in producing the correct running bias of the cell,
$B_{CCW}$, from the individual flagellar bias, $b_{CCW}$:
\begin{equation}
\label{eq:voting_hypothesis}
      B_{CCW} = \sum_{j=\nu}^{N_f}\left(\stackrel{N_f}{j} \right) \,
                 b_{CCW}^j \, (1-b_{CCW})^{N_f-j} \, ,
\end{equation}
where $\nu$ is the minimum number of flagella required 
to be in the CCW state for the cell to run.
Although cooperativity among the flagella is not yet fully understood, we similarly
adopt the ``majority rules'' algorithm to determine whether 
each cell will run or tumble in the presence of a chemoattractant gradient:
\begin{equation}
\label{eq:nu}
	\nu = \sum_{i=1}^{N_f} s^{(i)} \, .
\end{equation}
where $s^{(i)} = 0, 1$ is the state of the $i^{th}$ flagellum, corresponding to
CW and CCW directions of rotation, respectively.
Based on the experimentally measured gain of a single flagellum \cite{ref:Cluzel}, a steady
state value of $\bar{y}_p = 2.95 \, \mu {\rm M}$ yields $b_{CCW} \sim 0.64$
to within experimental error. The resulting running bias for the cell,
according to Eq.\ \ref{eq:voting_hypothesis}, is $B_{CCW} \sim 0.86$
for $N_f = 6$ and $\nu= 3$, in agreement with experimental results.  
Hence, if three or more flagella are determined to be in the CCW state, the
cell in our simulation runs; otherwise, it tumbles.

\subsection{Monte Carlo scheme}

For each cell, the Monte Carlo scheme can be outlined as follows.  At
$t_n$:
\begin{itemize}

\item[{(a)}] Each flagellum has a state $s_n$, 
where $s_n=0$ if CW, and $s_n=1$
if CCW.  To determine $s_{n+1}$, for each flagellum, 
we draw a uniformly distributed random number, $\zeta \in [0,1]$.

\item[{(b)}] We determine an approximation to the probability for
changing direction of rotation, given by Eq.\ \ref{eq:prob_switch}.  If the values
of $k_\pm$ are such that this approximation is not valid, $\Delta t$ is
reduced until it is.  This probability is a function of
[CheY-P], which depends on the trajectory that the cell has taken
along the external chemoattractant gradient.

\begin{itemize}
\item If the flagellum is in the CW state ($s_n=0$) and $\zeta < k_- \, \Delta t$,
it continues in the CW state ($s_{n+1}=0$); else, it switches to the CCW state ($s_{n+1}=1$).
\item If the flagellum is in the CCW state ($s_n=1$) and $\zeta < k_+ \, \Delta t$,
it continues in the CCW state ($s_{n+1}=1$); else, it switches to the CW state ($s_{n+1}=0$).
\end{itemize}

\item[{(c)}] If three or more flagella are now in the CCW state,
$\nu_{n+1} \ge 3$ in Eq.\ \ref{eq:nu}, and the cell runs; else, it tumbles.
If the cell is determined to run and $\nu_n < 3$, the direction of motion
is chosen to be left or right with equal probability.  Otherwise, it continues to run in
the same direction.

\item[{(d)}] The position of the $j^{th}$ cell, $x_n(j)$, 
is accordingly updated to $x_{n+1}(j)$ (using the
accepted time step $\Delta t$), with the cell
speed, $v_{cell}$, assumed to be constant and independent of position.

\item[{(e)}] The signal transduction variables $\left( u_1, u_2 \right)$ are
integrated in time, using the forward Euler scheme.  Their time history
is a function of each cell's trajectory.

\end{itemize}

\subsection{Combining signal transduction and motor response}

Although we use a toy excitation-adaptation model to describe signal transduction,
it is coupled to an experimentally realistic description of the flagellar switch.
To do so requires introducing (i) a shift in the steady state value of $u_1$, and
(ii) an amplification or gain factor, $g_0$, for the signal transduction network, required
so that the output of this network spans the dynamic range of the motor:
\begin{equation}
	y_p = \bar{y}_p - g_0 \, u_1
\end{equation}

The issue of gain in chemotaxis continues to receive 
much attention \cite{ref:amplification}.
Given the level of simplicity of our signal transduction model, we do not
imply this treatment of network gain to be physically realistic.  An obvious
drawback is that, like other parameters in the toy model 
(excitation and adaptation times), it is static,
whereas in the actual biological system, these parameters 
depend on the external signal. 

In Table\ \ref{table:parameters}, we report the numerical values for the
MC model parameters that operationally interface the toy signal transduction model
with the realistic model of the flagellar switch.  Hence, although
the choice of toy model parameters was ``physically motivated'', we do not attempt to 
make direct correspondence with experimental values.  We caution the reader
against making quantitative connection between our numerical
results at these values and experimental results.  


\begin{table}[htbp]
\caption{Numerical values for parameters used in the Monte Carlo evolution.}
\label{table:parameters}
\begin{center}
\begin{tabular*}{5.0in}{@{\extracolsep{\fill}}ll}
\hline
\hline
Parameter 	&Value \\
\hline
$\tau_a $  	&$100$ s \\
$\tau_e $	&$0.1$ s \\
$g_0 $		&$5$     \\
$N_T $		&$15 \,\, \mu {\rm M}$    \\
$K_L $	        &$1 \,\, \mu {\rm M}$ \\
\hline
$\bar{y}_p $     &$2.95 \,\, \mu {\rm M}$ \\
$K_M $	        &$3.1 \,\, \mu {\rm M}$ \\
\hline
$v_{cell} $      &$20 \, \, \mu {\rm m} \, \, {\rm s}^{-1}$ \\
\hline
\hline
\end{tabular*}
\end{center}
\end{table}


\section{Coarse integration}
\label{sec:ci}

For the present problem of simulating the chemotactic response of a population of
independent bacteria, the microscopic phase space is a directproduct
of the phase space for each cell, consisting of its
\begin{itemize}
	\item signal transduction variables: $\vec{u} = (u_1, u_2)$,
        \item flagellar state: $\vec{s}$, where $s^{(i)} = 0, 1$ for $i = 1 \ldots N_f$,
	\item position: $x$,
	\item direction of motion, $d = R, L, T$, corresponding to running right,
              running left and tumbling, respectively.
\end{itemize}
Rather than evolving all microscopic degrees of freedom, a coarse integration
scheme identifies suitable reduced variable(s) to be integrated.  
A clearly relevant reduced variable in population studies is the spatial
density distribution of cell positions,  $\mu_0(x,t)$, obtained from
the set of individual cell positions, $x_j$, which are a subset of the full 
microscopic phase space.  However, it is possible that the unknown
equations of the coarse description  use additional variables, for
example, the densities of the right moving, left moving, and
tumbling cells, $\rho_i$ where $i =  R, L,$ or $T$. In this case, 
$\mu_0(x,t) = \sum_i \rho_i(x,t)$.
Indeed, an optimal reduced representation balances accuracy and efficiency of
modeling the physical process.  Here, we propose to retain only the spatial
density distribution $\mu_0(x,t)$ as the relevant macroscopic variable in
coarse modeling of chemotaxis since, as we will demonstrate numerically, the density of
cells in each state rapidly approaches a functional of the total density.

A systematic approach to testing the adequacy of a particular coarse
description is discussed in \cite{ref:SGK}.  It is based on
locating the same fixed point at different levels of coarse
description, and examining the slow eigenvalues and corresponding
eigenvectors of the linearization of the coarse timestepper at
these fixed points (see also the discussion in \cite{ref:MMK}).

\begin{figure}[htbp]
   \begin{center}
   \includegraphics[width=4.5in]{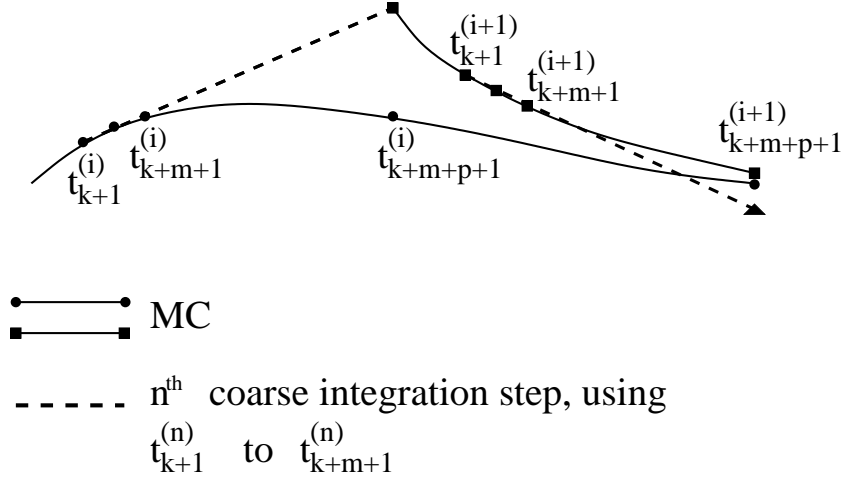}
   \end{center}
   \caption{
Schematic illustration of the coarse integration procedure.
}
\label{fig:sketch_mc}
\end{figure}

The separation of characteristic time scales describing the underlying
macroscopic and microscopic dynamics -- for example, the (long) time scale on which
the spatial density distribution of a population of bacteria moving in
a chemoattractant profile changes versus the (short) mean runtime of a single cell --
allows taking time steps in evolving the macroscopic
state that are ``long'' in comparison with the ``fast'' microscopic 
time scales in the problem, resulting in a computationally - 
efficient time evolution of the macroscopic state.  
We demonstrate this assumption of the separation of time scales explicitly
in Section\ \ref{sec:results}.
Figure\ \ref{fig:sketch_mc} shows a cartoon sketch of the 
\textsl{coarse integration} (CI) procedure, where the solid trajectories 
denote the \textsl{restriction} of the full dynamics onto a suitable low-dimensional subspace
(see Section\ \ref{sec:coarseint} for how a low dimensional representation of
the macroscopic spatial distribution, $\mu_0(x,t)$, is constructed).  
At each CI step, the coarse-integrated solution is \textsl{lifted} 
from the lower dimensional space into the full microscopic state space and 
evolved according to the Monte Carlo scheme, allowing
the error incurred during the coarse time step to relax to the slow
manifold parameterized by the cell density. Because the internal 
state $\{\vec{u}, \vec{s} \}$ of cells, and their
directions of motion, $d$, are ignored
in obtaining $\mu_0(x,t)$, they 
must be suitably reinitialized to construct the initial condition
for the next MC step (see Section \ref{sec:reinit}).

\subsection{SVD-based restriction scheme}
\label{sec:coarseint}

Below, we outline our coarse integration scheme which operates on
data from singular value decomposition of the distribution of cell
positions from Monte Carlo time evolution results.  
In the following, the solution at a ``reporting step'' 
refers to the sorted cell positions at specified 
time intervals, $T_{step}$.  Note that this time interval
should be distinguished from a single Monte Carlo
iteration, of which it contains a large number.  

\begin{enumerate}

\item The Monte Carlo description is simulated for
      $k$ reporting steps, corresponding to a total time interval 
      equal to $T_{settle}$.
      Step by step results are accumulated for a further $m$ 
      reporting steps, corresponding to 
      a total time interval equal to $T_{fit}$.

\item At each reporting step, we sort the cell positions, $x_i(j)$, to
      obtain $X_i(j)$, $i=k+1, \ldots, k+m$, and
      $j=1, \ldots, N_{\rm cells}$.
      The matrix $A_\ell$ is constructed (columnwise): $A_\ell = 
      \{\vec{X}_{k+1}, \vec{X}_{k+2}, \ldots, \vec{X}_{k+m}\}$.
      Figures\ \ref{fig:chemoattractant} and \ref{fig:sorted} show 
      the chemoattractant profile, and representative
      time sequences of sorted cell positions and
      corresponding histograms of the spatial density distributions, respectively.  
      However, we find it more convenient to work with the cumulative
      probability distribution function, defined as
\begin{equation}
	P(X,t) = \frac{1}{L} \, \int_{x_{min}}^X \mu_0(x^\prime,t) dx^\prime,
\end{equation}
which is related to the sorted cell positions at that time according to
\begin{equation}
	j/N_{cells} = P(X_i(j), t_i) \,,
\end{equation}
where $L \equiv x_{max} - x_{min}$ is the spatial domain.
We will use approximations to $P(X,t)$, or equivalently the
sorted cell positions, $\vec{X}_i$, 
as the macroscopic variable \cite{ref:Gear_distribution}.

\begin{figure}[htbp]
   \begin{center}
   \includegraphics[width=4.0in]{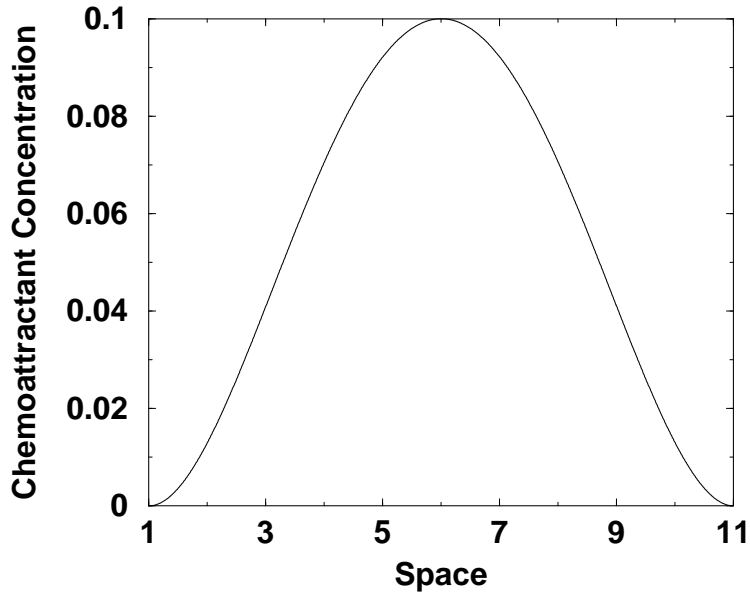}
   \end{center}
   \caption{
Chemoattractant profile used in Monte Carlo simulation of chemotaxis:
$S(x) = S_0 (x-x_{min})^2 (x-x_{max})^2$, where $S_0 = 1.6 \times 10^{-4}$ and 
$(x_{min}, x_{max}) = (1, 11)$.  The choice of this profile ensures
zero gradient of chemoattractant at the boundaries, consistent with no-flux boundary
conditions for the motion of bacteria.
}
\label{fig:chemoattractant}
\end{figure}

\begin{figure}[p]
   \begin{center}
   \includegraphics[width=4.0in]{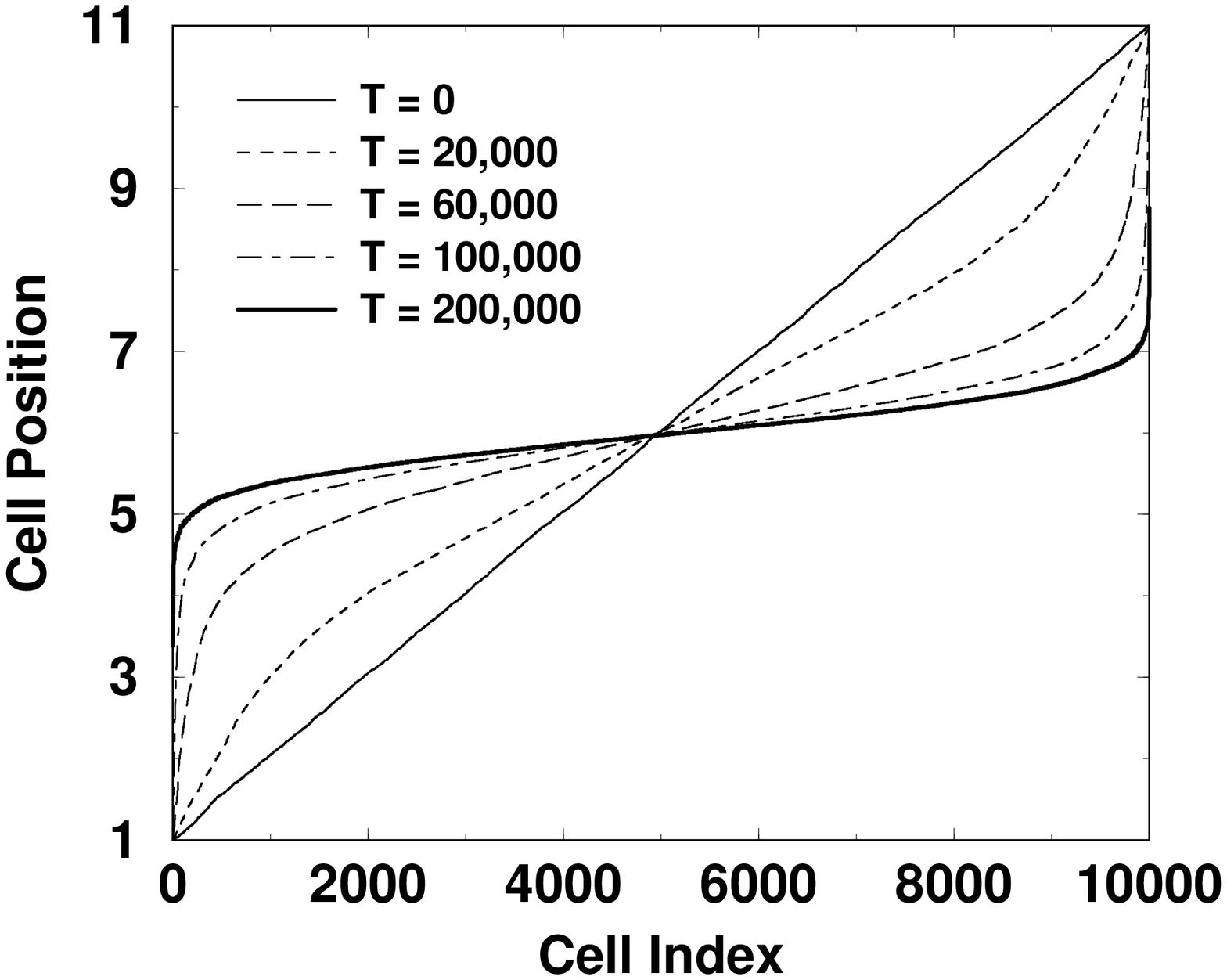} 
   \includegraphics[width=4.2in]{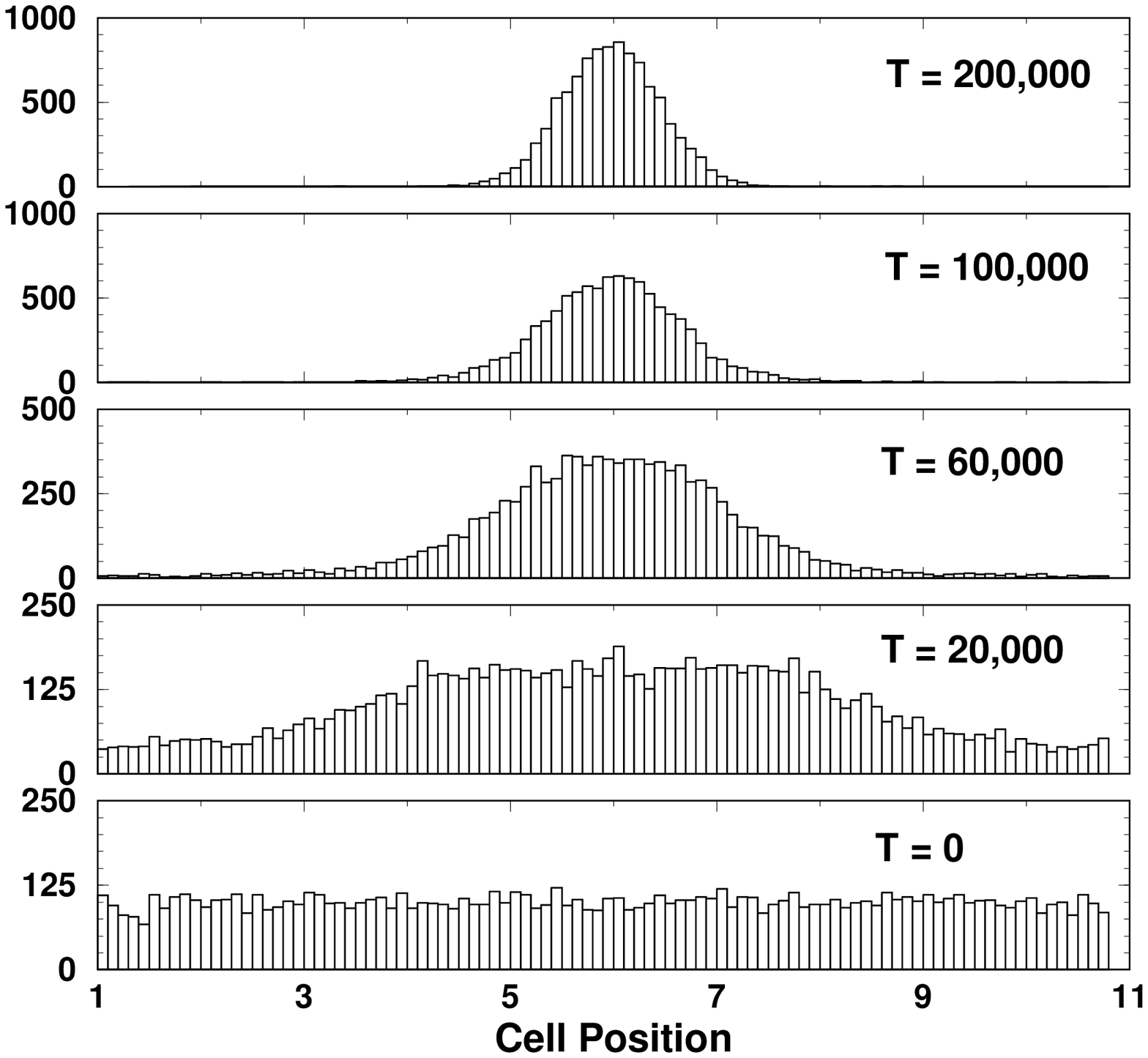}
   \end{center}
   \caption{
Sorted cell positions (top), and corresponding histograms (bottom) using a
bin size equal to $0.1$, at $T=0$ (thin solid line), $T=20,000$ 
(dashed line), $T=60,000$ (long-dashed line)
and $T=100,000$ (dot-dashed line), and $T=200,000$ (thick solid line).
From the time evolution of the variance of these distributions
we have determined that at $T=200,000$, the coarse solution
is very close to the stationary state.
}
\label{fig:sorted}
\end{figure}

\item A low dimensional representation of the sorted cell positions is 
      constructed in terms of orthonormal numerical basis functions, $\{ \vec{u}^{(r)} \}$, 
\begin{equation}
	\alpha^{(r)}(t_i) = \vec{u}^{(r)} \cdot \vec{X_i} = 
        \sum_{j=1}^{N_{cells}} u^{(r)}(j) \, X_i(j),
\end{equation}

where the basis functions are obtained from singular value decomposition of the MC data
as described below.

We distinguish between two approaches to obtaining the numerical
basis set. 

\begin{itemize}

	\item \textsl{Local basis functions:} 
              The singular value decomposition of $A_\ell$, constructed in Step 2,
              is computed:
	      \begin{equation} A_\ell = U_\ell \, W_\ell \, V_\ell^\top\end{equation}
              The columns of $U_\ell$, given by $\{ \vec{u_\ell}^{(r)} \}$, 
              $r=1,\ldots, m$, are the numerical basis functions.  We assume
              that they remain valid basis functions during the projected step.
              $W_\ell$ is a diagonal matrix, where 
	      $\left\{{w_\ell}_1, {w_\ell}_2, \ldots, {w_\ell}_m \right\}$
              are the singular values.

	\item \textsl{Global basis functions:}
              Using a single MC evolution from initial conditions to steady state, $T_f$,
              the matrix $A_g$ is constructed from the full data according
              to Step 2: 
              $A_g = \{\vec{X}_1, \vec{X}_2, \ldots, \vec{X}_{M}\}$,
              where $M = T_f/T_{step}$.  Singular value decomposition of $A_g$ gives the
              the global basis functions, $\{ \vec{u_g}^{(r)} \}$.
	      (In practice, initial data from 
              $t=0$ to $t=T_{settle}$ are not included.)
             
\end{itemize}

\item We perform linear least squares extrapolation of $\{ \alpha^{(r)}(t_i) \}$ using
      $i=k+1, \ldots, k+m+1$ to obtain
      $\{ \alpha^{(r)}(t_{k+m+1+p}) \}$,
      corresponding to a projected time equal to $T_{proj}$.  
      The projected solution is given by:
	\begin{equation} 
		\vec{Y}_{k+1+m+p} = \sum_r \alpha^{(r)}(t_{k+m+1+p}) \vec{u}^{(r)}   . 
        \end{equation}
\item In practice, we use only the numerical basis functions associated
      with the dominant singular values, $r=1, \ldots, r_{max}$, where the truncated
      number of basis functions, $r_{max}$, depends on the spectrum of singular
      values.

\end{enumerate}

The use of empirical orthogonal eigenfunctions, 
also referred to as the Karhunen-Lo\`eve
expansion or the Proper Orthogonal Decomposition, obtained
through singular value decomposition of experimental or simulation
data, for model reduction in systems with spatiotemporal dynamics
originates with Lorenz in the context of weather prediction
\cite{ref:Lorenz} and has found widespread use in the dynamical systems
context since the mid-80's (see the monograph \cite{ref:Holmes} and
references therein). 

The SVD process is computationally expensive, so if the computational
cost of the microscopic cell simulations were small, using global basis functions
generated in a preliminary run would be less costly than using local
basis functions.  However, in a simulation of a large number of 
cells, the total microscopic simulation cost for the cell population 
may be much larger that for the SVD process.  
For example, in the numerical experiments  reported here the CPU 
time for one simulation step of an individual cell was
2.07 microseconds, while the time for one SVD step was 22.4
milliseconds.  However, in a simulation of $10^4$ cells 
using local basis functions, the microscopic integration routine was called
$4\times 10^9$ times versus 16 calls on the SVD routine so that
0.0041\% of the time was spent on SVD.

\subsection{Reinitialization of internal variables}
\label{sec:reinit}

The question of consistent schemes for combining
microscopic and macroscropic descriptions of a physical
process is central to multiscale modeling.  
Here, to alternate each CI step with MC, in principle allowing relaxation
of the numerical error in the projective step, 
we must define an appropriate reinitialization
procedure: We know the position of each cell, $x$, after the CI
step, but we discard all information about its internal state,
$\{ \vec{u}, \vec{s} \}$, given by the values of the signal
transduction variables and the flagellar states, 
and its direction of motion, $d_j$.

\begin{figure}[htbp]
   \begin{center}
   \includegraphics[width=5.5in]{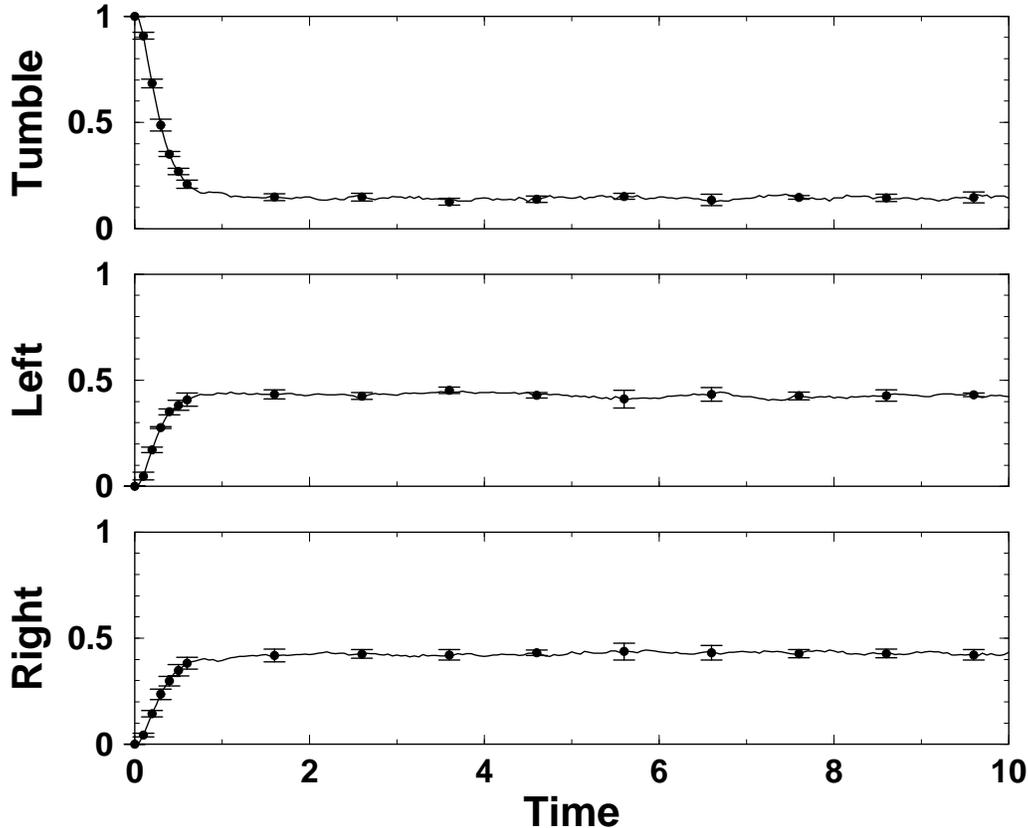}
   \end{center}
   \caption{
Ratios of right moving, left moving and tumbling cells at $x=4 \, (\pm 0.1)$, using
$j=1, \ldots, N_{cells} = 100,000$ and starting from an initially uniform distribution
of cells along the chemoattractant profile:  Averages over 5 Monte-Carlo realizations
started  from different random number seeds are plotted.  
The error bars, corresponding to the root-mean-square of the distribution of these ratios,
are shown for select points.
We note fast relaxation of
these ratios to their equilibrium values.
The signal transduction variables were initialized to their local equilibrium values,
$\vec{u}_j = \left\{0, f\left(S(x_j) \right) \right\}$. 
}
\label{fig:ratios}
\end{figure}

Our reinitialization protocol is empirically motivated:  the 
signal transduction variables are set equal to their local equilibrium values,
$\vec{u}_j = \left\{0, f\left(S(x_j) \right) \right\}$, where $j$ refers
to the $j^{th}$ cell.  This allows
the subsequent response of each cell to be within the most sensitive
range of the motor gain.  For simplicity, all (six)
flagella are restarted in the CW state,  $\vec{s}_j = \left\{0,0,\ldots,0\right\}$.
In Figure\ \ref{fig:ratios}, we show relaxation of the ratios of the numbers 
of right, left and tumbling cells at a set location along the chemoattractant profile.
These ratios at steady state for a flat chemoattractant profile are
\begin{equation}
	\frac{\rho_R+\rho_L}{\mu_0} = B,
\end{equation}
where the right and left moving ratios are equal.  For the present
choice of numerical parameters, $B \sim 0.86$ and $\rho_R = \rho_L \sim 0.43$.  
Other flagellar reinitialization schemes lead to similarly rapid relaxation
rates.

\section{Numerical results}
\label{sec:results}

\subsection{Low-dimensional representation}

\begin{figure}[htbp]
   \begin{center}
   \includegraphics[width=5.5in]{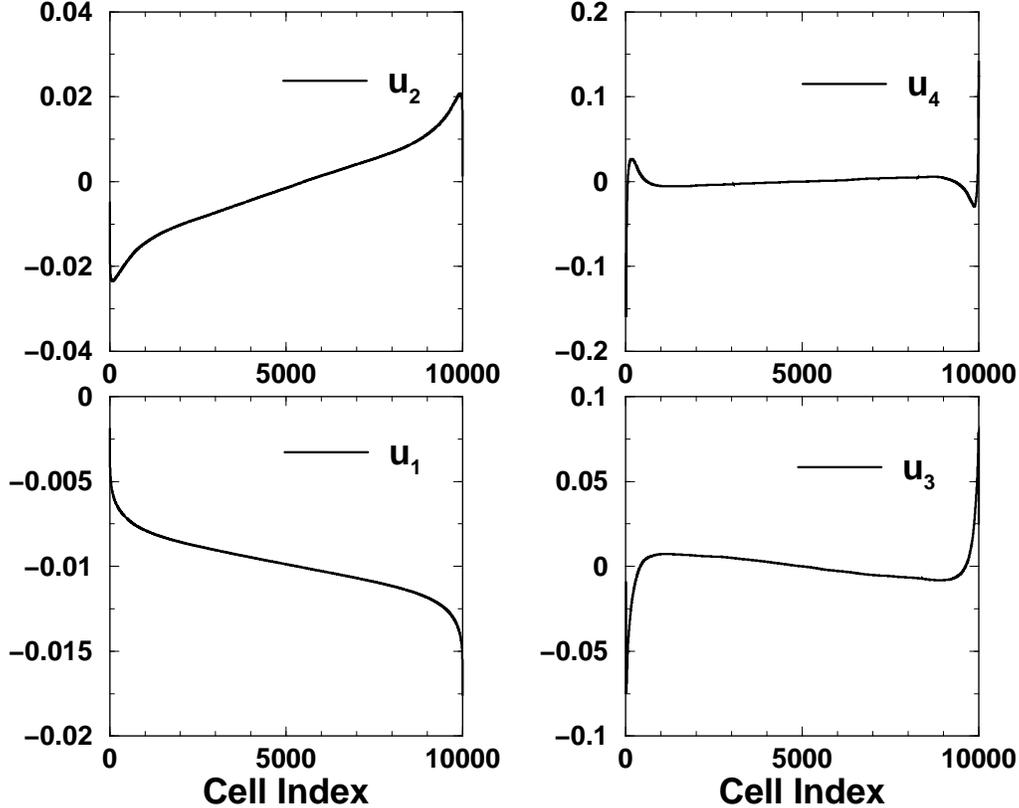}
   \end{center}
   \caption{
$\{ \vec{u}_g^{(r)} \}$, $r=1, \ldots, 4$ are the first four global SVD basis functions
for the cumulative cell density.
}
\label{fig:globalbasisfcns}
\end{figure}

\begin{figure}[htbp]
   \begin{center}
   \includegraphics[width=4.5in]{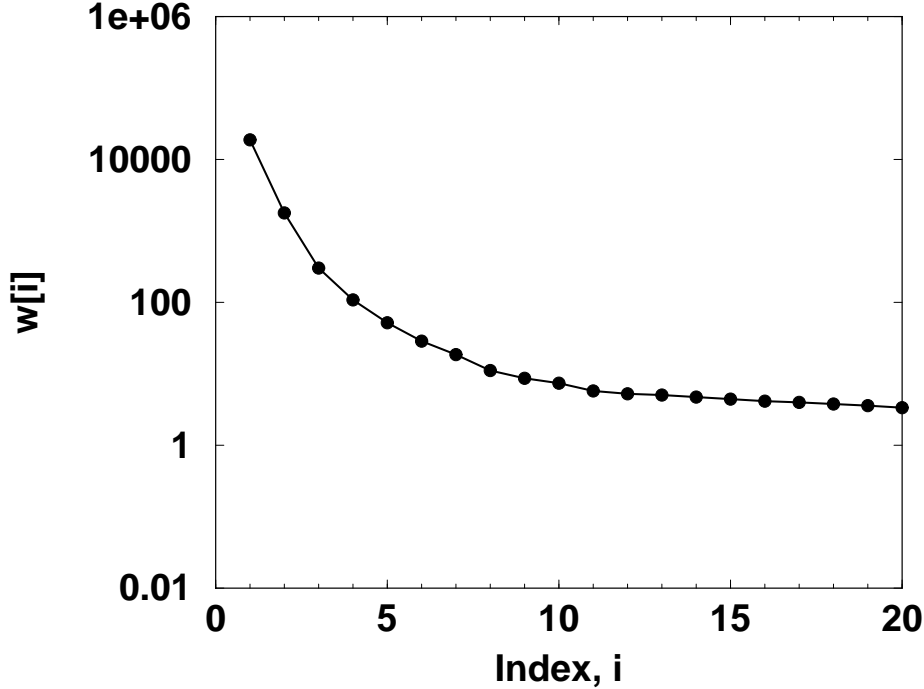}
   \end{center}
   \caption{
Largest singular values from global SVD of Monte Carlo data from initial
conditions to steady state ($T_f = 200,000, T_{step} = 200$).  The first
four dominant modes ($r_{max} = 4$) are used to construct the low-dimensional
representation of the cumulative cell density distribution.
}
\label{fig:singularvalues}
\end{figure}

\begin{figure}[htbp]
   \begin{center}
   \includegraphics[width=6.5in]{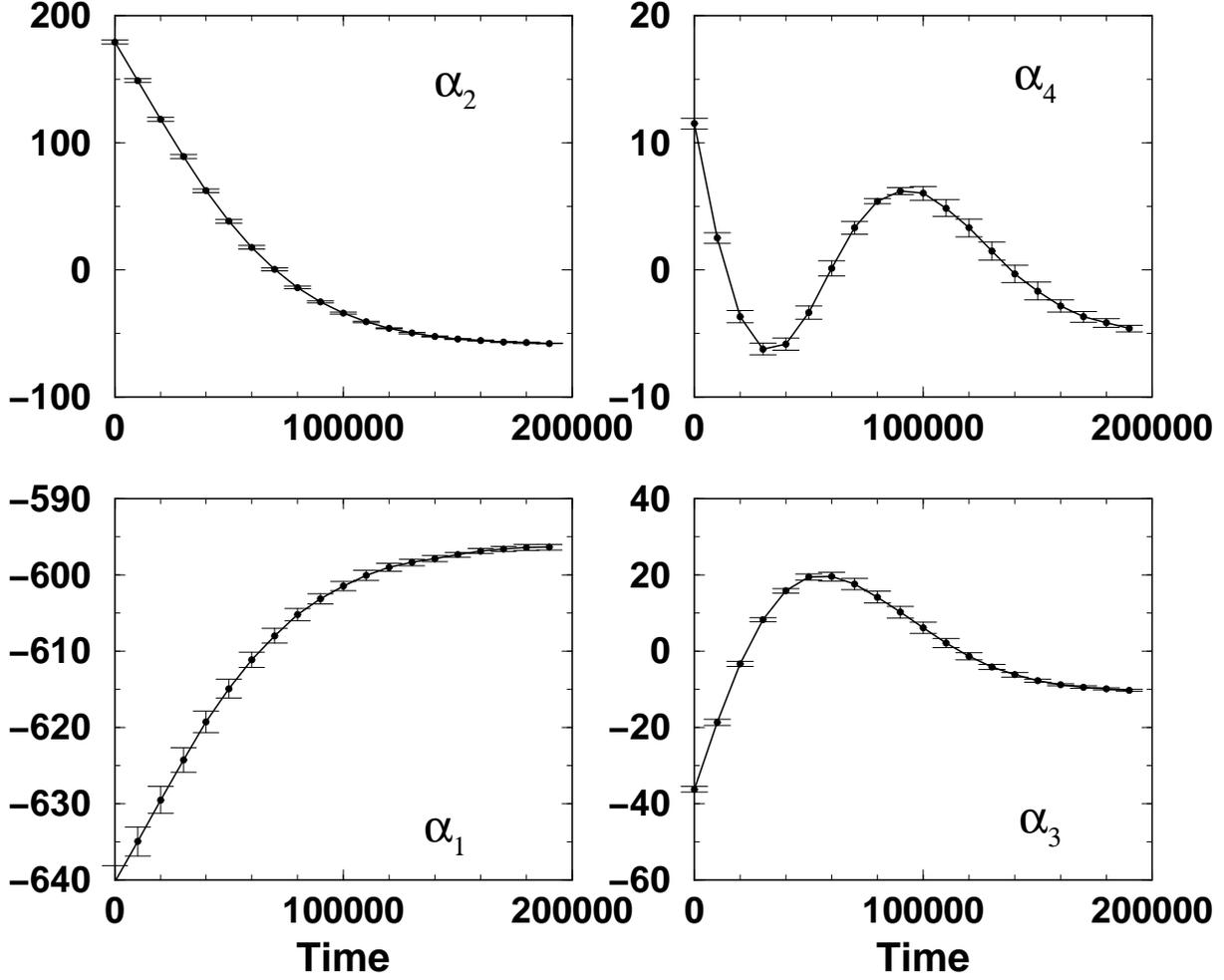}
   \end{center}
   \caption{
Coefficients of SVD basis functions:
$\alpha^{(r)}(t_i) = \vec{u}^{(r)} \cdot \vec{X_i} = 
\sum_{j=1}^{N_{cells}} u^{(r)}(j) X_i(j)$,
where the points represent averages over $N_{\rm MC}=10$ MC realizations, 
starting from different random
number seeds.  The global basis functions $\{ \vec{u}^{(r)} \}$ are obtained from one of the
MC runs.  The error bars, plotted for select points,
correspond to the root-mean-square of the distribution of
coefficients.
}
\label{fig:coeffs_av_mc}
\end{figure}

In Figures\ \ref{fig:globalbasisfcns} and\ \ref{fig:singularvalues}, 
we show the first four dominant global basis functions and singular values 
obtained from MC evolution of $N_{cells} = 10^4$ from initial conditions to
steady state.  Figure\ \ref{fig:coeffs_av_mc} shows the mean coefficients over $N_{\rm MC} = 10$
MC realizations
\begin{equation}
	\bar{\alpha}^{(r)}(t) = \frac{1}{N_{\rm MC}} \, \sum_{k=1}^{N_{\rm MC}} \alpha_k^{(r)}(t) \, ,
\end{equation}
and the root-mean-square of the distribution of coefficients, given by
\begin{equation}
\sigma_\alpha^{(r)}(t) = 
   \left\{ \frac{1}{N_{\rm MC}-1} \, \sum_{k=1}^{N_{\rm MC}} \left[ \alpha_k^{(r)}(t) - 
   \bar{\alpha}^{(r)}(t) \right]^2 \right\}^{1/2} \,.
\end{equation}
Hence, the error bars in this figure denote the expected statistical variation
in the values of these coefficients as a function of time.

\begin{figure}[htbp]
   \begin{center}
   \includegraphics[width=6.5in]{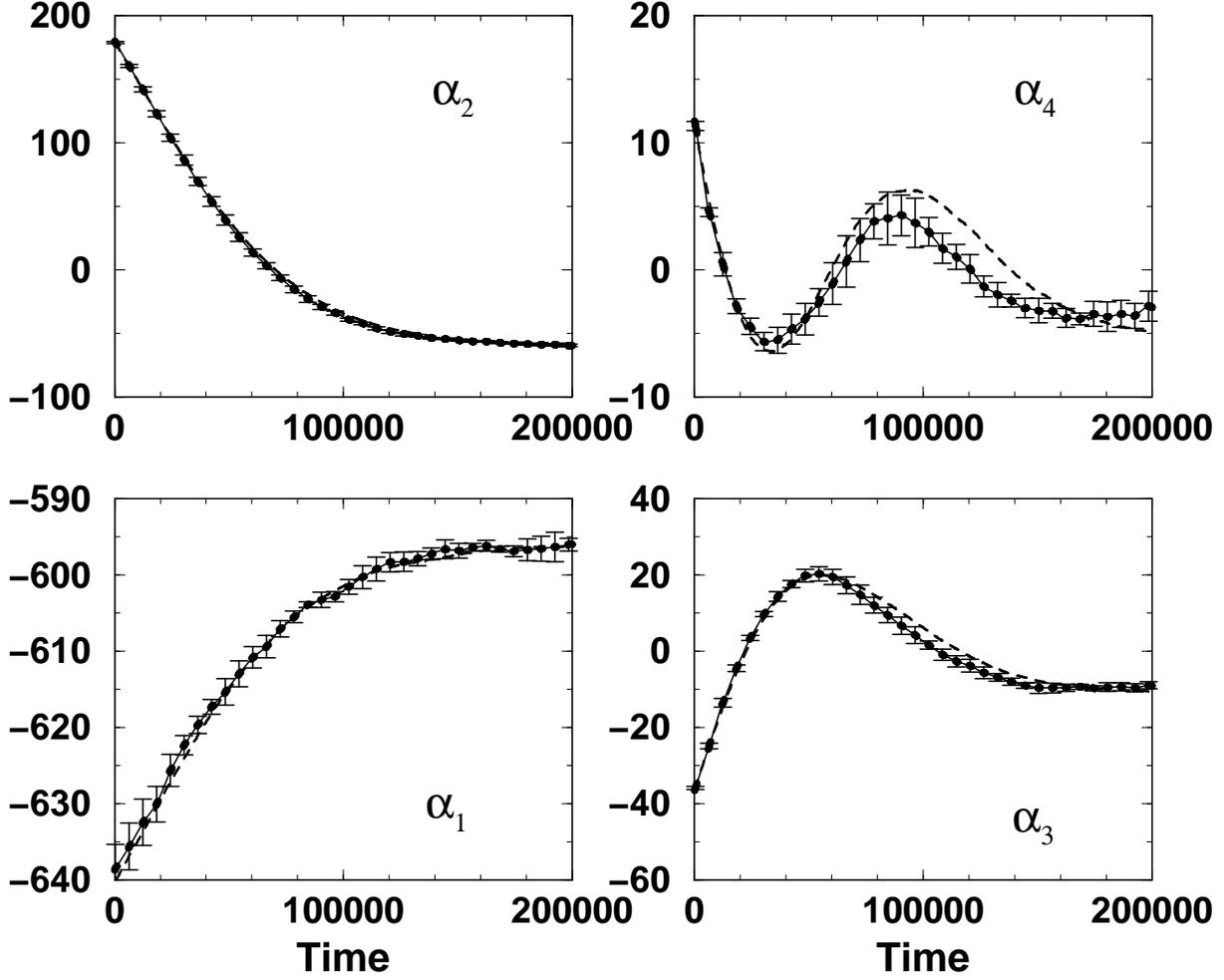}
   \end{center}
   \caption{
Coefficients of SVD basis functions:
$\alpha^{(r)}(t_i) = \vec{u}^{(r)}_g \cdot \vec{X_i} = 
\sum_{j=1}^{N_{cells}} u^{(r)}_g(j) X_i(j)$,
with CI parameters $\left( T_{settle}, T_{fit}, T_{proj} \right) =
10^3 \times (\frac{1}{2}, \frac{1}{2}, 5)$.  
The points represent averages over $N_{\rm MC}=5$ realizations, 
starting from different random number seeds.  
The error bars, plotted for select points,
correspond to the root-mean-square of the distribution of
coefficients. The average MC coefficients (dashed line) 
are shown for comparison.
}
\label{fig:coeffs_gsvd_500_500_5k}
\end{figure}

In Figure\ \ref{fig:coeffs_gsvd_500_500_5k}, we show the low-dimensional
representation of the coarse-integrated solution
for $T_{settle} = T_{fit} = 5 \times 10^2$ and $T_{proj} = 5 \times 10^3$.  For each
CI step, after evolving the Monte Carlo
for $T_{settle}$, we use
a linear fit to 
$\left \{\alpha^{(1)}(t), \alpha^{(2)}(t), \alpha^{(3)}(t), \alpha^{(4)}(t) \right\}$ 
during the interval $T_{fit}$ to project the solution forward in time by $T_{proj}$.
The points correspond to averages over $N_{\rm MC} = 5$ realizations, and the error bars
give the rms of the distribution of these coefficients.  For reference, we have included
the average MC coefficients in this figure.  These results indicate that for the higher order
coefficients, $\left\{\alpha^{(3)}(t), \alpha^{(4)}(t) \right\}$, 
whose coarse dynamics are described by a shorter characteristic
time-scale, a higher order time integration scheme would be more effective
than the explicit Euler method used here \cite{ref:RM}.  Consequently, although 
for intermediate times the difference between the CI and MC results
the straightforward CI scheme has captured the macroscopic
dynamics of the solution.

These results also illustrate the implicit assumption of separation of time scales:
The coefficients $\left\{\alpha^{(r)}(t) \right\}$,
governing the macroscopic behavior of the system,
vary on a time scale of ${\cal O}(10^4)$ units, 
while the longest microscopic time scale (adaptation time of
the signal transduction model) is ${\cal O}(10^2)$.

\subsection{Analysis of Errors}

For comparison of solutions at different values of the CI parameters,
$\left( T_{settle}, T_{fit}, T_{proj} \right)$, with the MC, we construct
the following measure of relative error:
\begin{equation}
	\varepsilon(t) = \left\{ \frac{\sum_{r=1}^4 \left[ \alpha^{(r)}(t) - 
                                                     \bar{\alpha}^{(r)}(t) \right]^2}
                                     {\sum_{r=1}^4 \left[ \bar{\alpha}^{(r)}(t) \right]^2}  
                                              \right \}^{1/2}.
\end{equation} 
The coefficients
$\{ \alpha^{(1)}(t), \alpha^{(2)}(t), \alpha^{(3)}(t), \alpha^{(4)}(t) \}$,
used in computing the relative error of the CI solution are
obtained as inner products of the solution with the global basis set,
regardless (a) of whether global or local basis sets were used in the
restricting/lifting step, and (b) the dimensionality of the restricted space.

\begin{figure}[htbp]
   \begin{center}
   \includegraphics[width=5.0in]{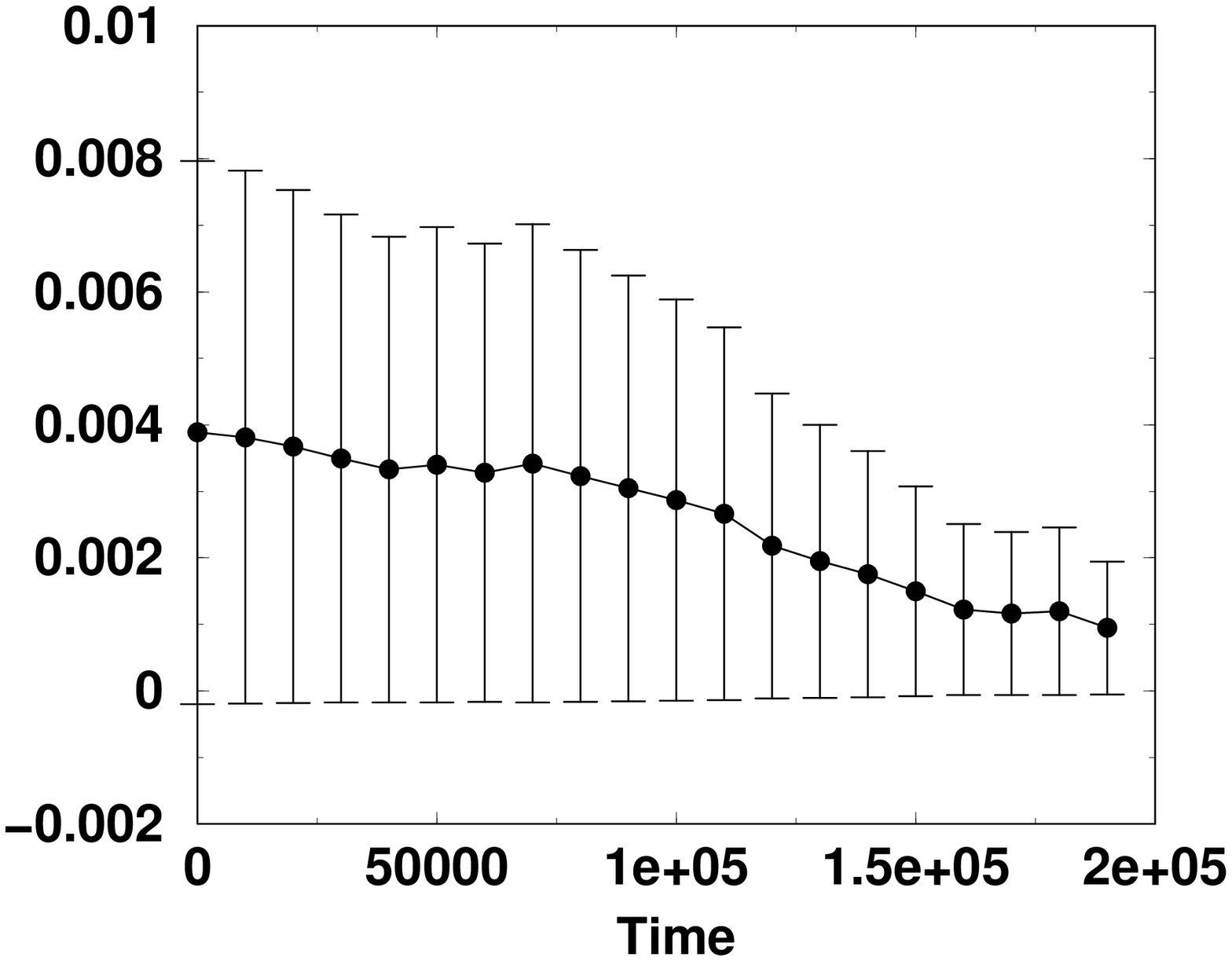}
   \end{center}
   \caption{
Relative error given by $\varepsilon_k(t) = \left\{ \sum_{r=1}^4 \left[ \alpha_k^{(r)}(t) - 
                                                     \bar{\alpha}^{(r)}(t) \right]^2 /
                                     \sum_{r=1}^4 \left[ \bar{\alpha}^{(r)}(t) \right]^2  
                                              \right \}^{1/2} $: 
The points correspond to the average error, obtained over $k=1, \ldots, 10$ 
MC realizations starting from different random number seeds.  
The error bars denote the error on the mean, $\sigma_{\bar{\varepsilon}}$. 
}
\label{fig:error_mc}
\end{figure}

In Figure\ \ref{fig:error_mc}, we first plot the 
average relative error for $N_{\rm MC}=10$ MC runs, given by:
\begin{equation}
	\bar{\varepsilon}(t) = \frac{1}{N_{\rm MC}} \, \sum_{k=1}^{N_{\rm MC}} \varepsilon_k(t).
\end{equation}
The error bars denote the error on the mean, given by 
$\sigma_{\bar{\varepsilon}} = \sigma_\varepsilon/\sqrt{N_{\rm MC}}$, where
$\sigma_\varepsilon$ is the root-mean-square of the
distribution of the error
\begin{equation}
	\sigma_\varepsilon(t) = \left \{ \frac{1}{N_{\rm MC}-1} \, \sum_{k=1}^{N_{\rm MC}} \left[
                            \varepsilon_k(t) - \bar{\varepsilon}(t) \right]^2 \right \}^{1/2}.
\end{equation}

\begin{figure}[htbp]
   \begin{center}
   \includegraphics[width=5.5in]{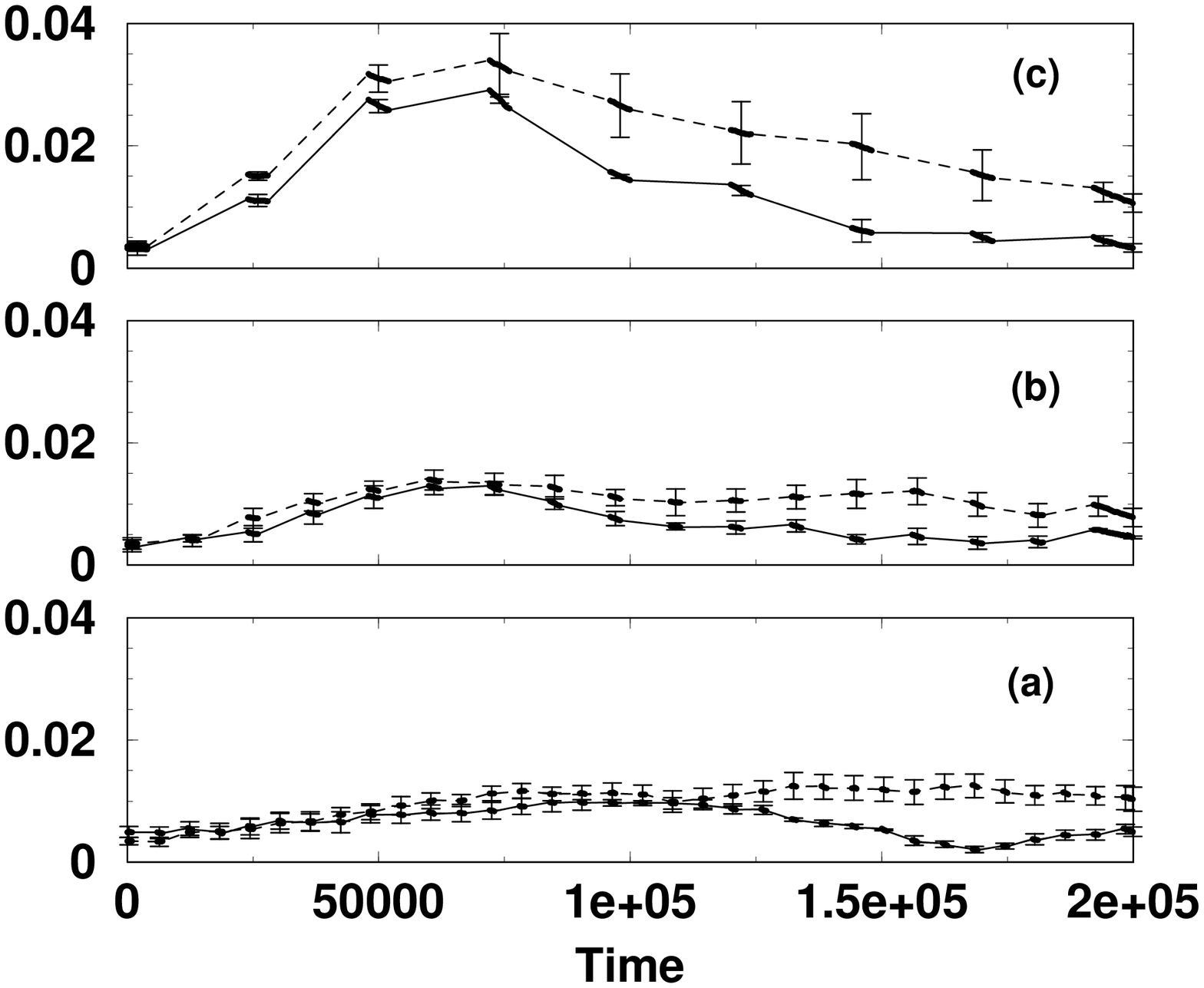}
   \end{center}
   \caption{
Relative error given by $\varepsilon(t) = \left\{ \sum_{r=1}^4 \left[\alpha^{(r)}(t) - 
                                                     \bar{\alpha}^{(r)}(t) \right]^2 /
                                     \sum_{r=1}^4 \left[ \bar{\alpha}^{(r)}(t) \right]^2 
                                              \right \}^{1/2} $,  
where $\alpha^{(r)}(t)$ are the coefficients of the global SVD basis functions
for CI solutions corresponding to parameters  
$\left( T_{settle}, T_{fit}, T_{proj} \right) =
10^3 \times (\frac{1}{2}, \frac{1}{2}, 5)$ (bottom), 
$(10^3, 10^3, 10^4)$ (middle), and 
$10^3 \times (2, 2, 20)$ (top).
The CI solution at each step was constructed 
using the first four global basis functions, shown using
solid lines, and using the first two local basis functions, shown using dashed lines.
The points correspond to averages over $N_{\rm MC} = 5$ realizations starting from
different random number seeds. 
The error bars denote the error on the mean, $\sigma_{\bar{\varepsilon}}$. 
}
\label{fig:error_dsvd_gsvd}
\end{figure}

We similarly compute the relative error for CI solutions, obtained
using either (i) global or (ii) local SVD basis functions in each step.
Figure\ \ref{fig:error_dsvd_gsvd} shows the mean 
errors for solutions with CI parameters given by
$\left( T_{settle}, T_{fit}, T_{proj} \right) =
10^3 \times (\frac{1}{2}, \frac{1}{2}, 5), 
10^3 \times (1, 1, 10)$ and 
$10^3 \times (2, 2, 20)$.
Note that for these parameter values,
the efficiency of the CI scheme remains the same.
In this figure, solid lines show results obtained using global SVD basis functions, 
$\left\{\vec{u}^{(1)}_g, \vec{u}^{(2)}_g, \vec{u}^{(3)}_g, \vec{u}^{(4)}_g \right\}$, 
in the CI steps.   Dashed lines show
these results obtained using local SVD basis functions. 
In the latter case, since singular value decomposition is applied to
a short segment of the MC, we found
the higher order local basis functions to be relatively
``noisy''; hence, $\left\{\vec{u}^{(1)}_l, \vec{u}^{(2)}_l \right\}$ 
were used in each CI step. Note that the
rms of the distribution of errors obtained for CI solutions using
local basis function is larger; in this case, the basis functions
themselves are also subject to statistical variations.
We find that the relative errors using global SVD basis 
functions are generally smaller than those corresponding to
CI solutions obtained using local SVD basis functions.
Using either global or local basis functions in the CI step,
the smallest relative error is achieved with the first two sets 
CI of parameters, corresponding to projected time intervals
short enough to resolve the macroscopic dynamics of
the higher order coefficients of the low-dimensional
representation of the solution.

\begin{figure}[htbp]
   \begin{center}
   \includegraphics[width=4.5in]{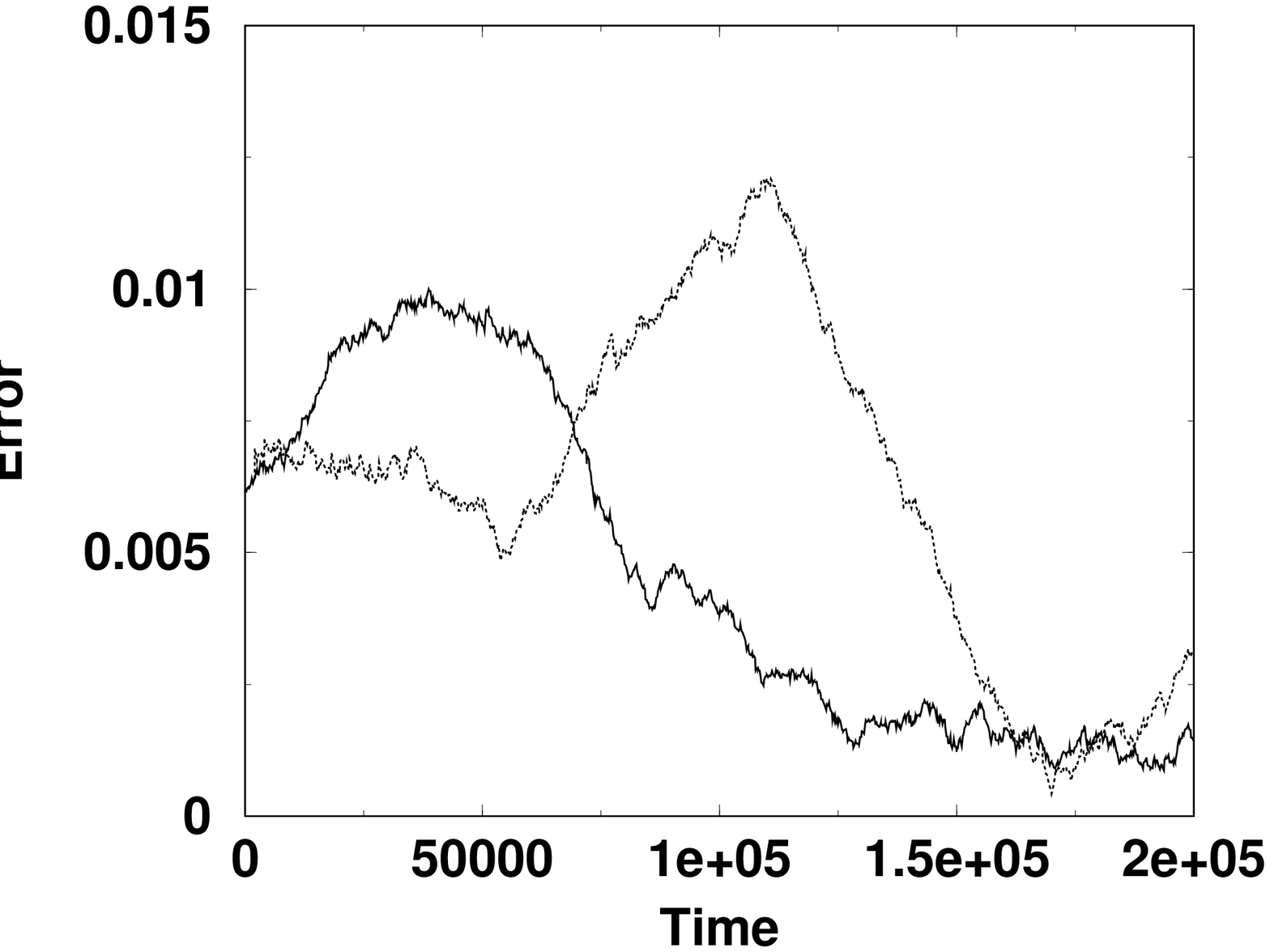}
   \end{center}
   \caption{
Relative error given by  $\varepsilon(t) = \left\{ \sum_{r=1}^4 \left[\alpha^{(r)}_g(t) - 
                                                     \bar{\alpha}^{(r)}(t) \right]^2 /
                                     \sum_{r=1}^4 \left[ \bar{\alpha}^{(r)}(t) \right]^2 
                                              \right \}^{1/2} $
with $\left( T_{settle}, T_{fit}, T_{proj} \right) =
10^3 \, \times \, (1, 1, 0)$:  After each $\left(T_{settle}+T_{fit} \right)$ interval,
the MC evolution is restarted from its low dimensional representation given by
$\left \{\alpha^{(1)}_g, \ldots, \alpha^{(4)}_g \right\}$ (dotted line) or
$\left \{\alpha^{(1)}_g, \ldots, \alpha^{(8)}_g \right\}$ (solid line).
}
\label{fig:error_noproj}
\end{figure}

As a benchmark, in Figure\ \ref{fig:error_noproj} we show the
error obtained for $\left( T_{settle}, T_{fit}, T_{proj} \right) =
10^3 \times (1, 1, 0)$, using global SVD functions $\{ \vec{u}_g^{(r)} \}$,
with $r=1, \ldots, 4$ (dotted line) and $r=1, \ldots, 8$ (solid line).
These results show the error incurred in restricting the spatial
distribution of cell positions to the low dimensional representation
given by $\{ \alpha^{(r)} \}$, without the contribution from the projective step.  
First, we note that the maximum error obtained from the
restriction to
$\{ \alpha^{(1)}_g, \ldots, \alpha^{(4)}_g \}$ and 
$\{ \alpha^{(1)}_g, \ldots, \alpha^{(8)}_g \}$
are comparable, indicating that using a higher dimensional restriction
is not significantly advantageous here.  
Secondly, this
error is comparable to the total error in Figure\ \ref{fig:error_dsvd_gsvd}(b),
which includes projective integration.  Given that the total error increases
with a longer projective time step, $T_{proj}$, 
as shown in Figure\ \ref{fig:error_dsvd_gsvd}(c),
these results point out that the ``optimal''
$T_{proj}$, for which the errors due to the projective time step and
restriction are separately comparable, is achieved 
in Figure\ \ref{fig:error_dsvd_gsvd}(b).

\begin{figure}[htbp]
   \begin{center}
   \includegraphics[width=4.5in]{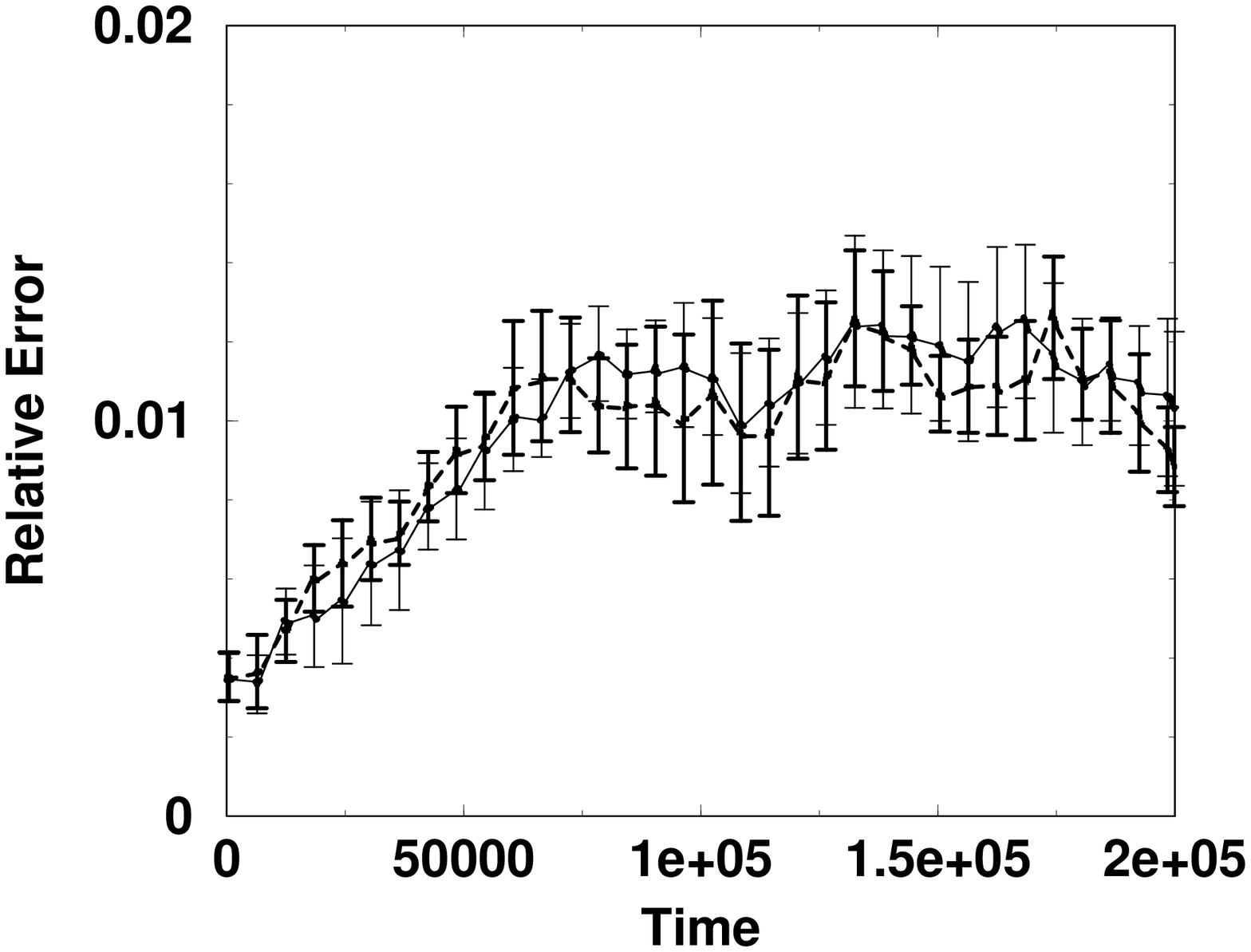}
   \end{center}
   \caption{
Comparison of relative errors for CI solutions (using local SVD basis functions)
corresponding to two different flagellar reinitialization schemes.
The signal transduction variables were initialized to their local equilibrium values,
$\vec{u}_j = \left\{0, f\left(S(x_j) \right) \right\}$.  In one case, 
all flagella were started in the CW state, shown in the thin solid line; 
in the second case, all flagella
were started in the CCW state, with those cells to the left (right) of the chemoattractant
peak running right (left), shown in the thick dashed line.
The points correspond to averages over $N_{\rm MC} = 5$ realizations starting from
different random number seeds. 
The error bars denote the error on the mean, $\sigma_{\bar{\varepsilon}}$. 
}
\label{fig:error_reinit}
\end{figure}

Finally, in Figure\ \ref{fig:error_reinit}, we compare CI errors using two different
reinitialization schemes and find them to be equivalent.

\section{Concluding remarks}
\label{sec:conclude}

We have demonstrated, through a coarse projective integration
algorithm, that short bursts of appropriately initialized microscopic
simulations can be used to simulate the macroscopic evolution of the cell 
density distribution in bacterial chemotaxis.
The outer level of our computational structure was a traditional,
continuum Euler integration scheme; the time-derivatives of the cell
density field it employed, however, were estimated from short
time evolution intervals of the Monte-Carlo description of this process, 
and not evaluated from a known, macroscopic expression.
This approach leads to a general framework for the computer-assisted
analysis of systems whose dynamics are given at a
microscopic/stochastic ``fine" level, but for which we require
averaged, macroscopic information.  

It is interesting  to discuss the benefits and shortcomings of
such a procedure. If accurate closed chemotactic
equations exist, one should use them instead of the two-tier modeling
we propose here.  In addition to the insight gained from exact or
approximate analytical solutions, computer-aided time-evolution or
bifurcation analysis using explicit equations is more efficient 
than kinetic Monte Carlo simulations.  
However, if such model equations are not available, our hybrid
computational approach can become more economical than long-time
direct Monte Carlo simulation.
Furthermore, when one is interested in qualitative transitions or 
bifurcations of the macroscopic behavior, marginally stable or unstable stationary
states may be difficult to identify through direct microscopic
simulations, while coarse timestepping holds promise when combined
with traditional bifurcation algorithms \cite{ref:MMK,ref:SGK,ref:RM}.  

It  appears, therefore,  that a modeler would ultimately 
gain in obtaining quantitative computational answers efficiently, but perhaps
lose in the fundamental understanding of a physical process 
that macroscopic model equations would offer.  Therefore a promising
research direction is to use such algorithms to test the 
validity of explicit closures that assume slaving of certain
higher order moments of the evolving distribution to lower order
ones.  In the chemotaxis context, when signal transduction and corresponding
motor response of the cell to an external signal are taken into account
as we do here,
macroscopic model equations derived systematically from
the microscopic description do not exist. 
The assumption implicit our the coarse integration algorithm 
was that the macroscopic description 
closes at the level of the spatial density distribution.  
In one spatial direction, for example, it would be interesting 
to learn when this assumption breaks down, requiring separate evolution of
right-going, left-going and tumbling cell density distributions.

More generally, we believe that the approach illustrated here for coarse time
integration of a Monte Carlo description of bacterial chemotaxis, is broadly applicable
with different types of microscopic simulators describing, for example, Brownian,
Lattice-Boltzmann, or molecular dynamics, and leading to emergent macroscopic dynamic behavior.
Such a hybrid scheme allows efficient simulation of the macroscopic behavior, and
may provide insight into macroscopic model equations.

\section*{ACKNOWLEDGEMENTS}
The work reported here was supported by the Princeton Council on Science and Technology
Fellowship (SS), NIH Grant \#GM29123 and
NSF Grant \#DMS 0074043 (HGO), and AFSOR (Dynamics and Control)
and NSF-ITR grants (CWG and IGK).


\end{document}